\documentclass[10pt,a4paper]{article} % 10pt is ignored!
\usepackage[dvips]{graphicx}
\usepackage{amsfonts,amsmath,amssymb}
\usepackage{hyperref}
\usepackage{cite}
\newcommand{\bse}{\begin{subequations}}
\newcommand{\ese}{\end{subequations}}

\begin{document}
\title{\bf Effect of electromagnetic fields on deformed $\mathrm{AdS}_5$ models}

\date{}
\maketitle
\vspace*{-0.3cm}
\begin{center}
{\bf Davoud Masoumi$^{a,1}$, Leila Shahkarami$^{b,2}$, Farid Charmchi$^{c,3}$}\\
\vspace*{0.3cm}
{\it {${}^a$Department of Physics, University of California, Merced, CA 95348, USA}}
\\
{\it {${}^b$School of Physics, Damghan University, Damghan, 41167-36716, Iran
}} \\
{\it {${}^c$School of Particles and Accelerators, Institute for Research in Fundamental Sciences (IPM),
P.O.Box 19395-5531, Tehran, Iran}}  \\
\vspace*{0.3cm}
{\it  {${}^1$dmasoumi@ucmerced.edu}, {${}^2$l.shahkarami@du.ac.ir}, {${}^3$charmchi@ipm.ir} }
\end{center}

\begin{abstract}
The response of a QCD-like gauge theory, holographically dual to a deformed $\mathrm{AdS}_5$ model, to constant electromagnetic fields is thoroughly investigated. 
The calculations in this paper are performed for three different cases, i.e., with only a quadratic correction, with only a logarithmic correction, and with both quadratic and logarithmic corrections, for which the parameters are chosen as the ones found in \cite{quadlog} by fitting to experimental and lattice results. 
The critical electric fields of the system are found by analyzing its total potential.
Comparing the total potential for the three cases, we observe that the quarks can be liberated easier in quadratic and then logarithmic case, for a given electric field.
Then, by calculating the expectation value of a circular Wilson loop, the pair production rate is evaluated while a constant electric field as well as a constant magnetic field are present.
The aforementioned result obtained from the potential analysis is also confirmed here when no magnetic fields are present.
We moreover find that the presence of a magnetic field perpendicular to the direction of the electric field suppresses the rate of producing the quark pairs and accordingly increases the critical electric field below which the Schwinger effect does not occur.
Interestingly, the presence of a parallel magnetic field alone does not change the response of the system to the external electric field, although it enhances the creation rate when a perpendicular magnetic field is also present.

\end{abstract}
Keywords: Schwinger effect; AdS/QCD; Confinement; Magnetic catalysis.
%$$$$$$$$$$$$$$$$$$$$$$$$$$$$$$$$$$$$$$$$$$$$$$$$$$$$$$$$$$$$$$$$$$$$$$$$$
\section{Introduction}
It is well known that the vacuum in quantum field theory (QFT) can be affected by external fields such as electromagnetic fields; virtual charged particle pairs become real if the external fields are strong enough. 
Such a phenomenon is referred to as the Schwinger effect \cite{Schwinger}.
Among other effects leading to the production of particles in high energy physics, such as the Hawking radiation from a black hole and the creation of particles in an expanding universe, the Schwinger effect is one of the most exciting ones due to its importance in understanding various aspects of QFTs.
The result that the dependence of the production rate of particles under the effect of a constant electric field is nonanalytic, shows its nonperturbative nature which motivates us to use it for exploring the nonperturbative regime of QFTs. 
This phenomenon can also provide an experimental tool for the verification of the results in the nonperturbative regime which has not been explored in a deep manner.

Although the Schwinger effect has not been observed directly yet by experiments, due to the very high intensity of the electric field, about $1.32\times 10^{18} Vm^{-1}$, needed for this effect to be detectable, new advances in laser technology have provided a promising ground to reach high-intensity electric fields and test the results of this phenomenon \cite{exper1,exper2}.

The Schwinger effect was first investigated at the weak coupling (in fact eliminating the Coloumb interaction) and weak field approximation \cite{Schwinger} and later generalized to the arbitrary coupling but weak field case\cite{manton1}.
Despite this effect has been first addressed in the context of quantum electrodynamics (QED), where the electron-positron pairs are created as a response to the electric field, it is also relevant to other QFTs, such as QCD.
Quark-antiquark pairs can be produced in the presence of electromagnetic fields, since they also have electric charges.
However, in this case, for the quark-antiquark pairs to be created, the electric force must overcome the confining force acting between the quarks.
Investigating this process, beside the fact that this is a realistic phenomenon and thus worth-noticing by itself, can be helpful in shedding light on the mysterious but not-yet discovered quark confinement.

One of the most important places that strong electromagnetic fields show up experimentally is in heavy-ion collisions, e.g., in RHIC and LHC, due to the highly accelerated charged particles \cite{sem1,sem2,sem3,sem4}.
In such strong electromagnetic fields, quark-antiquark pairs may be produced as well as electron-positron pairs.

Due to the nonperturbative nature of the Schwinger effect, it cannot be studied using the standard perturbation theory.
The AdS/CFT correspondence or generally gauge/gravity duality \cite{Maldacena2,Maldacena3,Maldacena1,Solana} provides a powerful framework to analyze this effect especially in the confining backgrounds such as QCD-like theories.
Following the work of Semenoff and Zarembo \cite{semenoff}, in which they proposed a holographic setup to investigate the Schwinger pair production rate in QED-like theories, a vast number of research has been done to explore the various aspects of this phenomenon in different systems with gravity duals including confining ones \cite{potential,Sch1,Sch2,Sch3,confin1,confin2,Sch5,Sch6,Sch7,confinrev,dehghani}.

AdS/QCD tries to construct a five-dimensional gravity theory in such a way that its dual field theory gains the known properties of the real QCD as much as possible.
Using this phenomenological approach, people employed some backgrounds which are reduced to the standard AdS in the UV but are different from it in the IR in such a way that the geometry is terminated at a finite value in the holographic radial direction.
The existence of the IR wall is crucial to model the confinement in any bottom-up holographic theory.
Such backgrounds that are shown to model experimental data or lattice results surprisingly well, can be simply constructed by slightly deformed $\mathrm{AdS}_5$ metrics.
It has been shown that these backgrounds have some important features in common with the QCD.
Most of them lead to the asymptotic linearity of Regge trajectories and a remarkable similarity to the Cornell potential for heavy-quarks, by fitting a small set of free parameters.
In \cite{quadlog} they employed such a background with a deformation function containing a quadratic and a logarithmic term with three free parameters.
They investigated the heavy-quark potential along with the dilaton field and dilation potential found from the Einstein equation and also examined the corresponding beta function for three cases: with only a quadratic correction, with only a logarithmic correction, and with both quadratic and logarithmic corrections.
The models with a negative and positive quadratic term known, respectively, as the soft wall \cite{soft} and Andreev-Zakharov \cite{andreev1,andreev2} model, have produced the properties related to the linear confinement including the linear Regge behavior of mesons in the former model and the linear Regge behavior along with the heavy-quark potential similar to the Cornell potential in the latter model.
Various aspects of the latter model have been considered in many other articles \cite{andreev3,light,lezgi}.
Using their investigations, the authors of \cite{quadlog} found the best values for the parameters of the model by the best fitted heavy-quark potential and other quantities mentioned above, in any of the three cases and in general they found the theory with only logarithmic correction to be fitted better than other theories.

%One of the most interesting features of SU(N) gauge theories is that they undergo a phase transition to a deconfined phase at high temperature. Only at very high temperatures far enough from the critical temperature can the perturbation theory be used to explore the properties of the system. However, some difficulties arise when considering the system near the phase transition point which comes from the non-perturbative effects. AdS/CFT correspondence builds a very powerful framework to resolve this problem. %Its spectrum is like that of the linear Regge models and therefore they were able to fix the value of the free parameter of the metric, which is the cause of the deformation from AdS, to fit the $\rho$ meson trajectory. This fitting leads to the approximate value $c\approx 0.9 GeV^2$.

Although a large number of papers have investigated different aspects of the Schwinger effect in different systems including confined ones, many other aspects are yet to be known, especially the ones regarding the effect of the simultaneous presence of electric and magnetic fields.
In this regard, our plan is to consider the response of the same theory as the one in \cite{quadlog} to an external electromagnetic  field.
Since the theory is confining and has an IR cutoff which is a generic feature of confining theories, we expect the existence of two critical electric fields, usually denoted by $E_s$ and $E_c$.
Below $E_s$, the Schwinger effect does not happen at all.
For electric fields between $E_s$ and $E_c$ the pairs are produced with an exponential suppression.
Above $E_c$ the pairs are produced freely and catastrophically and the vacuum of the theory is completely unstable.
The organization of the paper is as follows.
In the next section we set the basic framework that we are working in. 
Then, the total potential of quarks in the presence of an external electric field is evaluated in Sec. 3 and the radius of the IR wall in the three cases of our interest along with their critical electric fields are also calculated.
Section 4 is devoted to the calculation and analysis of the pair production rate by extremizing the world-sheet action of a string ending on the boundary of a circular Wilson loop located on a probe D3-brane in the bulk, and equipped with a constant electric and magnetic field.
In this section we concentrate mainly on the response of the system to the magnetic fields imposed in different directions.
We finally summarize our results in Sec. 5.
%$$$$$$$$$$$$$$$$$$$$$$$$$$$$$$$$$$$$$$$$$$$$$$$$$$$$$$$$$$$$$$$$$$$$$
\section{The deformed $\mathrm{AdS}_5$ model}
We shall begin by introducing the following Euclidean background metric:
\begin{align}\label{metric}
  ds_5^2=G_{m n}^s dX^m dX^{n}=e^{2{\cal A}_s(z)}\left(dt^2+d\vec{x}^2+dz^2\right),
\end{align}
where $G_{m n}^s$ denotes the metric in the string frame.
The radial direction is denoted by $z$ and the boundary is located at $z=0$.
%A deformed warp factor has been added to the pure $\mathrm{AdS}_5$ to break the conformal symmetry and find QCD-like models in the dual gauge theory side.
To break the conformal symmetry and find QCD-like models in the dual gauge theory side, a deformed warp factor has been added to the pure $\mathrm{AdS}_5$.
We define the deformation function $h(z)$ using $e^{2{\cal A}_s(z)}=\frac{h(z)L^2}{z^2}$, in which $L$ is the $\mathrm{AdS}_5$ radius.
$h(z)=1$ gives the Euclidean pure $\mathrm{AdS}_5$ metric. 
Following \cite{quadlog} we choose the deformation function in the form of
\begin{align}\label{metricf}
h(z)=\exp \left[\frac{\sigma z^2}{2}+\lambda \ln \left(\frac{z_{IR}-z}{z_{IR}}\right)\right],
\end{align}
with three free parameters. $\sigma$ and $\lambda$ can be either positive or negative and $z_{IR}>0$. If $\lambda=0$, the positive and negative values of the parameter $\sigma$ correspond to the Andreev-Zakharov model \cite{andreev1,andreev2} and soft-wall model \cite{soft}, respectively.

In what follows, we want to examine the effect of a static external electric field on this background for three cases; with only the quadratic function, with only the logarithmic term, and with both quadratic and logarithmic terms.
%$$$$$$$$$$$$$$$$$$$$$$$$$$$$$$$$$$$$$$$$$$$$$$$$$$$$$$$$$$$$$$$$$$$$$
\section{Studying the Schwinger effect using the total potential}
This section is devoted to the analysis of the total potential of a quark-antiquark pair  influencing by a constant external electric field $E$, in the background of our interest.

We know that the potential of a quark-antiquark pair of infinite masses is evaluated from the Wilson loop of a rectangle ${\cal C}$ with one direction along the time direction ${\cal T}$ and the other along the separation direction of the two quarks. 
For infinite ${\cal T}$ the expectation value of the Wilson loop gives the interaction potential of the heavy quark-antiquark pair.
Holographically, this can be found by calculating the on-shell Nambo-Goto (NG) action of an open string hanging down from the boundary with its endpoints separated by a distance $x$ in, say, $x_1$ direction of the field theory.
Since the creation of infinitely heavy quark particles is severely suppressed, to study the Schwinger effect, we should assume quarks of finite mass. 
To that purpose, following the Semenoff and Zarembo's proposal \cite{semenoff}, we put a probe D3-brane at an intermediate position $z_0$ in the radial direction and attach the endpoints of the string to this brane.
 
Parametrizing the bulk coordinates as ($t=\eta_0,~x_1=\eta_1,~x_2=x_3=0,~z=z(\eta_1)$) and obtaining the induced metric on the string world-sheet, the string action is given by
\begin{align}\label{ngaction}
S_{\mathrm{NG}}=\frac{1}{2 \pi \sigma_s}\int d\eta_0 d\eta_1 e^{2{\cal A}_s(z)}\sqrt{1+z'^2}=\frac{{\cal T}}{2 \pi \sigma_s}\int_{-x/2}^{x/2} d x_1 e^{2{\cal A}_s(z)}\sqrt{1+z'^2},
\end{align}
where $\sigma_s$ is proportional to the inverse of the string tension with a dimension of $GeV^{-2}$.
The conserved quantity is found to be $\frac{e^{2{\cal A}_s(z)}}{\sqrt{1+z'^2}}=e^{2{\cal A}_s(z_c)}$, where the right-hand side has been determined by the use of the conditions $z(0)=z_c$ and $z'(0)=0$. %is the radial value at the tip of the string, at which
By virtue of this relation and then substituting it into Eq.\,(\ref{ngaction}), the distance between the pair of quarks and their potential energy are derived respectively as follows:
\begin{align}\label{distance}
x=2 z_0 a\int_{1/a}^1 dv \frac{v^2\frac{h(z_0 a)}{h(z_0 a v)}}{\sqrt{1-v^4\left(\frac{h(z_0 a)}{h(z_0 a v)}\right)^2}},
\end{align}
\begin{align}\label{qpotential}
V_{Q \bar{Q}}=\frac{ L^2}{\pi \sigma_s  z_0 a}\int_{1/a}^1 dv \frac{h(z_0 a v)}{v^2\sqrt{1-v^4\left(\frac{h(z_0 a)}{h(z_0 a v)}\right)^2}},
\end{align}
in which we have defined the rescaled dimensionless quantities $v\equiv z/z_c$ and $a\equiv z_c/z_0$.
The relation (\ref{qpotential}) contains both the static (mass) and potential energy of the quark-antiquark pair. 
To obtain the total energy in the presence of an external electric field $E$, we need to add the potential associated with $E$, which leads to 
\begin{align}\label{vtotal}
V_{\mathrm{tot}}(x)=\frac{L^2}{\pi \sigma_s z_0 a}\int_{1/a}^1 dv \frac{h(z_0 a v)}{v^2\sqrt{1-v^4\left(\frac{h(z_0 a)}{h(z_0 a v)}\right)^2}}-E x,
\end{align}
where $x$ is the separation length of the quarks given in Eq.\,(\ref{distance}).
%===================================================
\subsection{IR cutoff of the deformed $\mathrm{AdS}_5$}
In almost all holographic confining gauge theories, there exists a wall cutting off the AdS space in the IR region, realizing the linear confinement.
One of the first efforts in this direction was led to the hard wall model \cite{hard}.
In the theories we are considering here also there is an IR cutoff $z_h$ in the gravity side, i.e., the radial coordinate $z$ is only allowed in the interval $z_0\leqslant z<z_h$, where $z_0$ is the position of the probe D3-brane.

The value of $z_h$ can be simply found by imposing the reality condition of the integral in Eq.\,(\ref{distance}) giving the separation length between the quarks in the dual gauge theory. 
In fact this condition should be satisfied in order for the chosen geometry could be the holographic dual of the gauge theory.
Using a similar argument presented in \cite{andreev2} for the case of the positive quadratic function (the Andreev-Zakharov model), one can simply obtain the maximum value of the radial position $z$ that the tip of a string hung from the boundary (or the D3-brane) can reach, which is equal to the IR cutoff of the gravity theory.
A brief explanation of the process of obtaining $z_h$ is presented here.
We suppose that the probe D3-brane is absent for a moment to make the argument simpler without changing the result.
This is equivalent to $z_0=0$ in our calculations.
To ensure that the integrand in this equation is real in the whole integral range 
$0\leqslant v \leqslant 1$, the function 
$f(v)\equiv 1-v^4\left(\frac{h(z_c)}{h(z_c v)}\right)^2$ in the denominator should be positive for the whole $v$-interval, where we have used the replacement $z_c=z_0 a$. 
A simple analysis shows that in all three cases of our interest this function is equal to one at the beginning of the integral range, $v=0$, it is zero at $v=1$, and tends to one again as $v \to +\infty$. 
One concludes immediately that this function has at least one minimum at which $f(v)\leqslant 0$.
To ensure the positivity of the function $f(v)$ for the whole integral range $0\leqslant v \leqslant 1$, we demand this minimum happens at $v_{\mathrm{min}}\geqslant 1$.
When $v_{\mathrm{min}}= 1$, the integral develops a logarithmic singularity and for $v_{\mathrm{min}}> 1$, it is definitely real.
By the use of the condition $v_{\mathrm{min}}\geqslant 1$, we find a constraint on the permitted values of $z_c$ in terms of the parameters of the theory as $z_c \leqslant z_h$.
 
Here we report the results, found by this calculation, for the three cases of our interest.
In the quadratic case where $\lambda=0$, the function $f(v)$ can be written as
\begin{equation*} 
f(v)=1-v^4 e^{\sigma z_c^2 \left(1- v^2 \right)}. 
\end{equation*}
The only extremum of this function found with the aid of the condition $\frac{d f(v)}{d v}=0$ is obtained as $v_{\mathrm{min}}=\sqrt{\frac{2}{\sigma z_c^2}}$. 
Then from $v_{\mathrm{min}}\geqslant 1$, we obtain $z_c \leqslant \sqrt{\frac{2}{\sigma}}$ meaning that $z_h=\sqrt{\frac{2}{\sigma}}$.
For the logarithmic case, i.e., $\sigma=0$, we can write 
\begin{equation*} 
f(v)=1-v^4 e^{-2 \lambda \log \frac{z_{IR}-z_c v}{z_{IR}-z_c}}=1-v^4 \left( \frac{z_{IR}-z_c v}{z_{IR}-z_c}\right)^{-2 \lambda},
\end{equation*}
and following the above mentioned procedure  we simply find $z_h=\frac{2 z_{IR}}{2-\lambda}$.
And, for the last case where both the quadratic and logarithmic terms are present the function $f(v)$ can be written as follows
\begin{equation*}
f(v)=1-v^4 \exp\left[{\sigma z_c^2 \left(1-v^2\right)-2 \lambda \log \frac{z_{IR}-z_c v}{z_{IR}-z_c}}\right].
\end{equation*}
This function has only one real extremum from which we can again obtain the IR cutoff in terms of the parameters of the theory.
In this case from the condition $\frac{d f(v)}{d v}=0$, we arrive at the algebraic equation $\sigma z_c^3 v^3 -\sigma z_{IR} z_c^2  v^2 +(\lambda-2) z_c v+2 z_{IR}=0$.
Using the condition $v_{\mathrm{min}}\geqslant 1$ in which $v_{\mathrm{min}}$ is the real root of this algebraic equation, one finds the value of $z_h$ in terms of the parameters of the theory.

Notice that $a=\frac{z_c}{z_0}$ which is the rescaled radial value of the tip of the string in the bulk cannot be larger than $\frac{z_h}{z_0}$.
In fact, at the maximum value of $z_c$, i.e., $z_h$ a logarithmic singularity is developed in the integral of Eq.\,(\ref{distance}) and the effective string tension reaches its minimum.
Also, the integral becomes complex for $z_c>z_h$.
%Increasing the value of the deformation parameter $c$ from zero leads to deviation from $\mathrm{AdS}_5$ black hole metric.
%This metric behaves as Euclidean $\mathrm{AdS}_5$ at $r\to \infty$, for any value of the parameter $c$.
%===================================================
\subsection{Critical electric fields}
According to the previous studies, for every confining field theory, there are two values of electric fields at which the response of the theory alters critically.
One of them, usually represented as $E_s$ and restricted to confined theories, is the threshold value of the electric field required for liberating the lightest quarks and starting the Schwinger effect.
The other critical electric field, usually denoted as $E_c$, is the one above which there is no potential barrier and the vacuum is completely unstable. 
The existence of $E_c$ is common in deconfined as well as confined phases.
For $E_s<E<E_c$ the quarks are faced with a finite potential barrier and they can be liberated only through a tunneling process.

In this section we search for the critical electric fields in the confined theories of our interest.
As mentioned above, $E_c$ is the electric field at which the total potential barrier  vanishes completely or equivalently $\lim_{x\to 0} \frac{d V_{\mathrm{tot}}}{d x}= 0$.
To obtain the derivative of the total potential with respect to $x$, we can use the chain rule for the first term in Eq.\,(\ref{vtotal}) as $\frac{d V_{Q \bar{Q}}}{d x}=\frac{d V_{Q \bar{Q}}}{d a}/\frac{d x}{d a}$.
From Eq.\,(\ref{distance}), it can be seen that $x\to 0$ is equivalent to $a \to 1$, where $a$ is related to the maximum radial value of the string. 
The derivatives are obtained as
\begin{align}\label{vprime}
\frac{d V_{Q \bar{Q}}}{d a}&=\frac{- L^2}{\pi \sigma_s a^2 z_0}\int_{1/a}^1 dv \frac{h(z_0 a v)}{v^2\sqrt{1-v^4\left(\frac{h(z_0 a)}{h(z_0 a v)}\right)^2}}+\frac{ L^2}{\pi \sigma_s a z_0}\int_{1/a}^1 dv \frac{\partial}{\partial a}\frac{h(z_0 a v)}{v^2\sqrt{1-v^4\left(\frac{h(z_0 a)}{h(z_0 a v)}\right)^2}}\nonumber\\&+\frac{L^2}{\pi \sigma_s a z_0}\frac{h(z_0)}{\sqrt{1-\frac{1}{a^4}\left(\frac{h(z_0 a)}{h(z_0)}\right)^2}},\\\label{xprime}
\frac{d x}{d a}=&2 z_0\int_{1/a}^1 dv \frac{v^2\frac{h(z_0 a)}{h(z_0 a v)}}{\sqrt{1-v^4\left(\frac{h(z_0 a)}{h(z_0 a v)}\right)^2}}+2 a z_0\int_{1/a}^1 dv \frac{\partial}{\partial a}\frac{v^2\frac{h(z_0 a)}{h(z_0 a v)}}{\sqrt{1-v^4\left(\frac{h(z_0 a)}{h(z_0 a v)}\right)^2}}\nonumber\\&+2 a z_0\frac{\frac{h(z_0 a)}{h(z_0)}}{\sqrt{1-\frac{1}{a^4}\left(\frac{h(z_0 a)}{h(z_0)}\right)^2}}.
\end{align}
All the integrals in both Eqs.\,(\ref{vprime}) and (\ref{xprime}) vanish in the limit $a\to 1$. 
Thus, we simply have
\begin{align}\label{vtox}
\frac{d V_{Q \bar{Q}}}{d x}=\frac{L^2}{2 \pi \sigma_s a^2 z_0^2}\frac{h\left(z_0\right)^2}{h(z_0 a)}.
\end{align}
Since $\lim_{a\to 1} h(z_0 a)=h(z_0)$, we finally arrive at
\begin{align}\label{limit}
\frac{d V_{\mathrm{tot}}}{d x}=\frac{L^2}{2 \pi \sigma_s}\frac{h(z_0)}{z_0^2}-E,
\end{align}
which gives the critical electric field as $E_c=\frac{1}{2 \pi \sigma_s}\frac{L^2 h(z_0)}{z_0^2}$.

The critical electric field $E_c$ can also be evaluated from the DBI action of a probe D3-brane including a constant world-volume electric field, located at the radial position $z_0$ in the bulk.
To do so, we need to work in the Lorentzian signature in which the first term in the deformed $\mathrm{AdS}_5$ metric (\ref{metric}) is negative. 
The DBI action simply reads
\begin{align}\label{dbiD3}
S_{\mathrm{D3}}&=-T_{\mathrm{D3}}\int d^4 x \sqrt{-\det \left(g_{\mu \nu} +2 \pi \sigma_s F_{\mu \nu}\right)}\nonumber\\
&=-T_{\mathrm{D3}}\int d^4 x \frac{h\left(z_0\right)L^2}{z_0^2}\sqrt{\left(\frac{h\left(z_0\right)L^2}{z_0^2}\right)^2-\left(2 \pi \sigma_s\right)^2 E^2},
\end{align}
where $T_{\mathrm{D3}}$ is the D3-brane tension.
The integral is only real for $E\leqslant E_c$ where $E_c=\frac{1}{2 \pi \sigma_s}\frac{L^2 h\left(z_0\right)}{z_0^2}$ which agrees with the critical value obtained from the analysis of the total potential of the quarks.
Notice that $E_c$ is a function of $z_0$, meaning that the critical electric field above which the pairs can be produced freely depends on the mass of the quarks, since according to the AdS/CFT dictionary the quark mass corresponds to the self-energy of a straight string stretched from the probe D3-brane, placed at $z_0$, to the IR-cutoff at $z_h$, which can be written as $m=\frac{L^2}{2\pi \sigma_s}\int_{z_0}^{z_h}\frac{h(z) dz}{z^2}$ in our problem.
Thus, by changing the position of the probe D3-brane $z_0$, the mass of the corresponding quark would change.

%Let us now consider the other critical electric field, i.e., $E_s$ above which the Schwinger effect begins to occur.
Now, we turn our attention to the other critical electric field, $E_s$, below which quark-antiquark pairs even the ones of zero mass cannot be created from the vacuum.
Thus, $E_s$ is the electric field at which the quarks are faced with an infinitely large  potential barrier, i.e., $\lim_{x\to \infty} \frac{d V_{\mathrm{tot}}}{d x}= 0$, or equivalently the total potential becomes completely flat.

Performing the same calculations as those illustrated above, the critical electric field is found as $E_s=\frac{1}{2 \pi \sigma_s}\frac{L^2 h\left(z_h\right)}{z_h^2}$.
Using a simple comparison, one can see that the relations of both of the critical electric fields are consistent with the ones found for the general backgrounds in \cite{Sch1}.
%===================================================
\subsection{Potential analysis}
Let us now study the effect of the external electric field on the theory of our interest.
In all the following results we use the values found and reported in \cite{quadlog}.
We set the AdS radius as $L=1~GeV^{-1}$, and the string tension as $\sigma_s=0.38~GeV^{-2}$ in all cases.
Also, for the quadratic deformation function we choose $\sigma=0.22~GeV^{2}$ and for the logarithmic deformation function the related parameters are chosen to be $\lambda=-0.272$ and $z_{IR}=2.1~GeV^{-1}$.
And, for the case with both the quadratic and logarithmic terms the parameters are $\sigma=-0.34~GeV^{2}$, $\lambda=-1$ and $z_{IR}=2.54~GeV^{-1}$.
Using these values, the critical electric field below which the pair production does not occur  is $E_s\approx 0.125234~GeV^{2}$, $E_s\approx 0.218321~GeV^{2}$ and $E_s\approx 0.248432~GeV^{2}$ for the three theories mentioned above, respectively. 
The other critical electric field $E_c$ depends on the mass of the quarks through $z_0$, the position of the probe D3-brane.
In all the calculations we set $z_0=0.5~GeV^{-1}$.

\begin{figure}[h]
\begin{center}
\includegraphics[width=6.8cm]{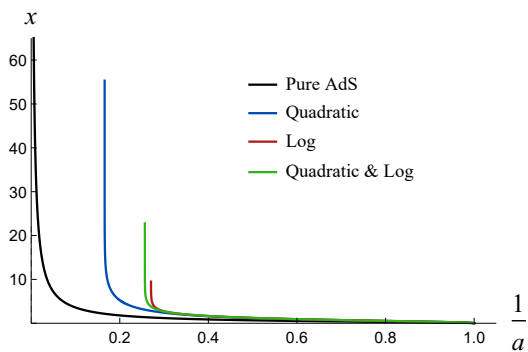}
\end{center}
\caption{\footnotesize 
The separation length of the quarks versus the rescaled radial position of the turning point of the string for different theories.}
\label{xa}
\end{figure}

We first depict the graphs of the separation of the quark-antiquark pair, given in Eq.\,(\ref{distance}), as a function of $1/a$.
Here, $1/a$ is the inverse of the rescaled radial position $z$ of the tip of the corresponding string in the bulk, which takes its values in the interval $[z_0/z_h,1]$, where $z_h$ is the maximum value of $z$ that can be reached in any case.
For the pure AdS, $z_h \to \infty$, i.e., the whole radial interval is attainable.
The separation lengths of the quarks for different cases are shown in Fig.\,\ref{xa}.
By increasing the distance $x$ between the quarks, the tip of the corresponding string moves farther away from the boundary until it lies on a horizon at $z=z_h$, where the quarks are eventually infinitely far from each other.
This happens at a finite value of $z_h$ for all the deformed AdS cases and is a characteristic of any confining field theory.
However, in the pure AdS case, as can be seen, the quarks can be infinitely far away from each other only when the tip of the string has reached the end of the AdS space at $z\to \infty$.

\begin{figure}[h]
\begin{center}
\includegraphics[width=6.8cm]{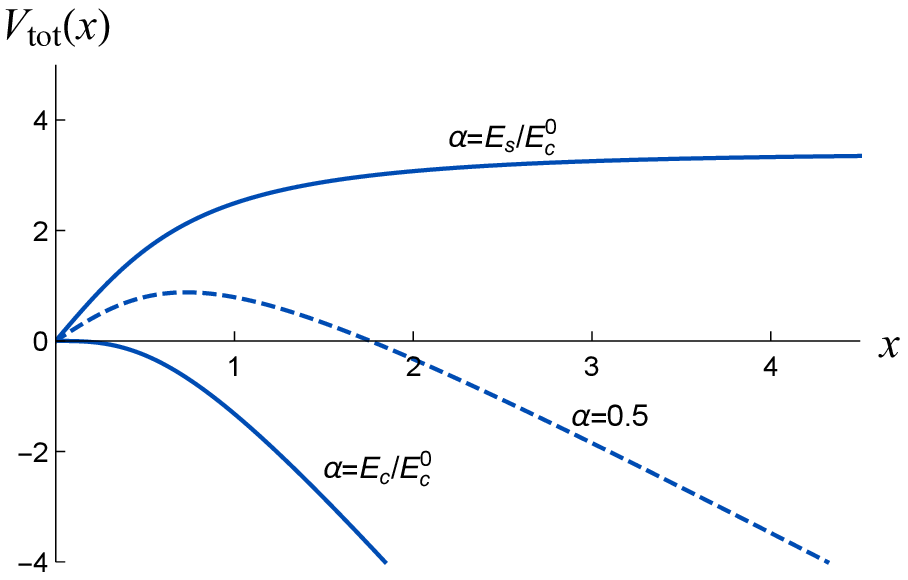}\hspace{.3cm}
\includegraphics[width=6.8cm]{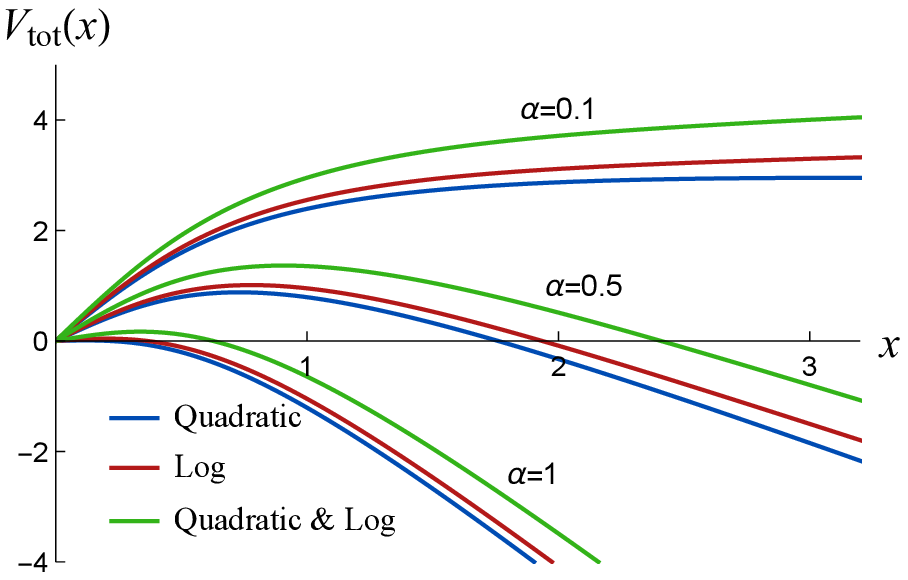}
\end{center}
\caption{\footnotesize 
Left graph: The total potential versus $x$ for various electric fields in the case of the quadratic deformation function.
Right graph: Comparison of the total potential for different deformation functions.}
\label{vtot}
\end{figure}

The results for the total potential of the quarks in the presence of the electric field are summarized in Fig.\,\ref{vtot}.
Notice that in this figure and in the following results the parameter $\alpha \equiv \frac{E}{E_{c}^0}$ denotes the rescaled electric field with respect to the critical electric field of the pure AdS case, i.e., $E_{c}^0=\frac{1}{2 \pi \sigma_s}\frac{L^2}{z_0^2}$.
The left panel shows the total potential for the case with only a quadratic deformation function.
As is obvious, the slope of the diagram at $x=0$ goes to zero for $E=E_c$ and the diagram of $E=E_s$ becomes flat as $x \to \infty$, as expected from the discussions and calculations of the previous subsection.
Moreover, for a sample electric field between these two critical fields, the quarks are faced with a finite potential barrier. 
In the right panel we compare the total potential for three values of the electric field and for three different deformation functions.
It can be seen that, $\mathrm{PB}_{\mathrm{Quad}}<\mathrm{PB}_{\mathrm{Log}}<\mathrm{PB}_{\mathrm{Quad \& Log}}$, where PB stands for the potential barrier.
This means that the quarks can be liberated simpler and faster in the quadratic case than in other cases.
Since the Schwinger effect happens through a tunneling process, a smaller potential  barrier means that the quarks can be freed for a smaller value of the electric field.

Notice that the choice of the form of the deformation function in the gravity side, per se, does not have any meaning in the physics of the field theory side.
In fact, in the bottom-up holographic theories one pursues the geometries with free parameters that lead to the most similar results to the experimental and lattice data.
And, at the present, there exists no real experimental data in this subject as a reference.
However, among these geometries with the parameters found in \cite{quadlog}, the one leading to the results with the most similarity to the upcoming results by the next generation light sources aiming at detecting the Schwinger effect is favored as a good holographic model of QCD.

We should mention a very important issue here.
The difference of the total potential of the three theories for a given external electric field comes from the fact that we have used the parameters found in \cite{quadlog} using the best fitted {\it heavy}-quark potential of these theories to the Cornell potential with the coefficients adjusted to fit the charmonium spectrum.
Notice that in our problem the mass of the quarks is chosen to be finite.
This assumption is required in order to be able to study the effect of the external electromagnetic fields on the system.
Therefore, we use the parameters found in \cite{quadlog} as the only available values, which would work as an approximation for our case with a different quark mass.
%*********************************************************************
\section{Pair production rate}
This section is devoted to the calculation of the pair production rate $\Gamma$.
This quantity is equivalent to the expectation value of a circular Wilson loop, in the $t-x_1$ plane, on the probe D3-brane located at $z_0$. 
$x_1$ is the direction of the applied electric field.
Holographically, we need to evaluate the NG action of a string, attaching to the D3-brane, coupled to a constant electric NS-NS 2-form $B_2=B_{01}dx^0\wedge dx^1$.
Then, according to the AdS/CFT dictionary, $\Gamma$ corresponds to the exponential of the total action as follows
\begin{align}
\Gamma \sim e^{-S}=e^{-S_{\mathrm{NG}}-S_{B_2}}.
\end{align}
The problem we are dealing with is to consider the effect of both electric and magnetic fields on the confining theory of our interest.  
In \cite{Sch2} the proposal of Semenoff and Zarembo was generalized so as to contain the study of the pair production rate in the presence of both electric and magnetic fields.
We first describe the setup to calculate $\Gamma$ in the presence of a constant electric field and then study the creation of quarks under the influence of both electric and magnetic fields.

To calculate the extremal surface in the bulk that shares the same boundary as the temporal-spatial Wilson loop on the D3-brane, we first need to obtain the induced metric on the string world-sheet.
We choose the following ansatz for the bulk coordinates
\begin{align}
t=r(\rho) \cos (\phi), ~~x^1=r(\rho) \sin (\phi),~~ z=z(\rho),
\end{align}
where ($\rho,\phi$) are the polar coordinates of the string world-sheet, and $0 \leqslant \phi \leqslant 2 \pi$ and $0 \leqslant \rho \leqslant \rho_0$.
The other coordinates of the bulk are chosen to be zero.
Having found the induced metric on the string world-sheet using the above ansatz and inserting it into the NG action, we arrive at
\begin{align}\label{ng1}
S_{\mathrm{NG}}=&\frac{L^2}{\sigma_s} \int_0^R d r \frac{r\ h(z)}{z^2}\sqrt{1+z'^2},\\\label{ng2}
S_{B_2}=& -\frac{1}{\sigma_s} B_{01}\int_0^R dr \ r= -\frac{R^2}{2\sigma_s} B_{01},
\end{align}
where $R$ is the radius of the circular Wilson loop on the D3-brane.
Notice that we have converted the integration variable from $\rho$ to $r$, and supposed that $r(\rho=0)=0$ and $r(\rho=\rho_0)=R$.
When only an electric field is turned on, the only nonvanishing component of $B_2$ is $B_{01}=E \left(2 \pi \sigma_s \right)$. 
From Eq.\,(\ref{ng1}), the equation of motion for $z(r)$ is given by
\begin{align}
2 r \left(1+z'^2\right)+z\left[z'+z'^3-r \frac{1}{h(z)}\frac{d h(z)}{dz}\left(1+z'^2\right)+r z''\right]=0,
\end{align}
where a prime denotes the derivative with respect to $r$.
This equation is solved numerically with the boundary conditions $z'(0)=0$ and $z(0)=z_c$ with $z_0 \leqslant z_c \leqslant z_h$.
Then, to find the solution consistent with the presence of $B_2$, we select the configuration that satisfies the following constraint condition 
\begin{align}
z'|_{z_0}=-\sqrt{\frac{1}{\alpha^2}-1}.
\end{align}
Through this constraint, classical action depends on the value of the electric field.

\begin{figure}[h]
\begin{center}
\includegraphics[width=6.8cm]{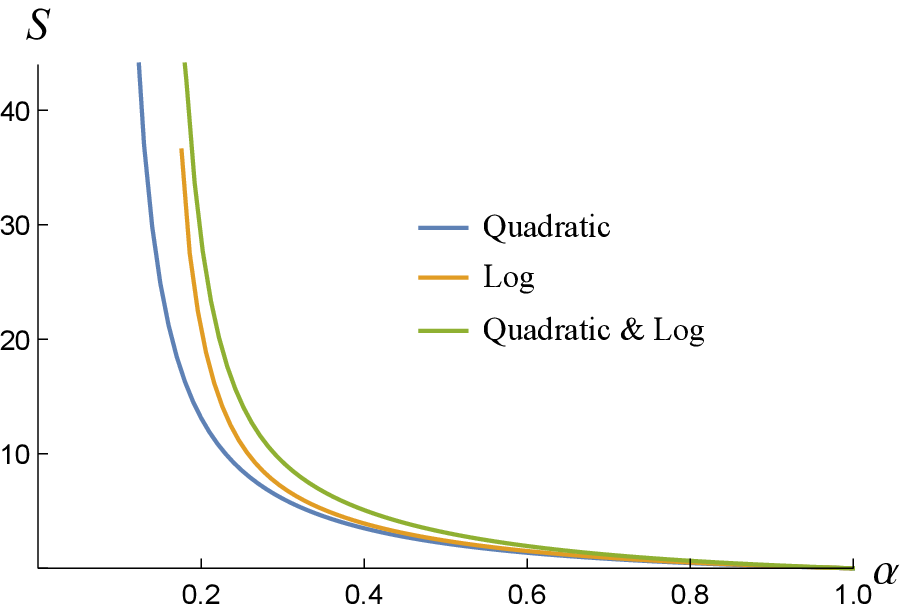}\hspace{.3cm}
\includegraphics[width=6.8cm]{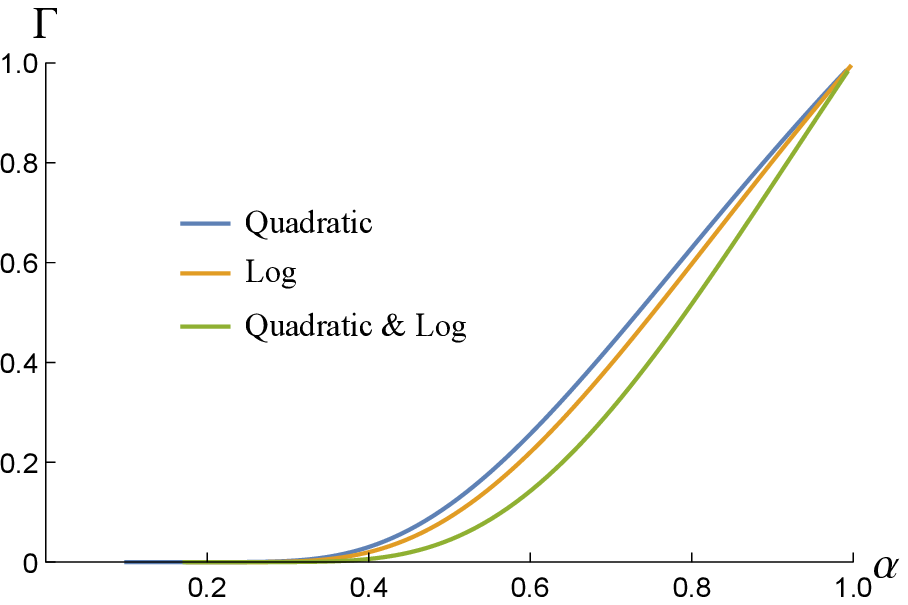}
\end{center}
\caption{\footnotesize 
The left and right graphs respectively show the action and the production rate $\Gamma$ versus the rescaled electric field for different deformation functions.}
\label{decayrate}
\end{figure}

The total action and its exponential part, i.e., the decay rate $\Gamma$ are respectively  shown in the left and right graphs of Fig.\,\ref{decayrate}, where different lines indicate the results for different deformation functions with the same parameters as in the previous section.
As can be easily seen, for an arbitrary value of the electric field $\Gamma_{\mathrm{Quad}}>\Gamma_{\mathrm{Log}}>\Gamma_{\mathrm{Quad \& Log}}$.
This means that the pairs are produced easier in the case with a quadratic deformation function and in the case with both quadratic and logarithmic functions a stronger electric field is needed to have the same pair production rate as the other two cases.
However, when the electric field approaches the critical electric field $E_{c}^0$, the difference between the production rates in various cases becomes less obvious. 

Let us now turn to the case where both electric and magnetic fields are present.
Using the results of \cite{Sch2}, the critical electric field $E_c^B$ in the presence of parallel and perpendicular magnetic fields can be found, from analyzing the DBI action of the probe D3-brane placed at $z_0$, as follows
\begin{align}\label{ecb}
E_c^B=E_c \sqrt{1+\frac{B_{\perp}^2}{E_c^2+B_{\parallel}^2}},
\end{align}
in which $E_c$ is the critical electric field in the absence of magnetic fields, and $B_{\perp}$ and $B_{\parallel}$ are the components of the magnetic field in the perpendicular and parallel directions with respect to the direction of the electric field, respectively.
It can be seen that the critical electric field is independent of $B_{\parallel}$ when $B_{\perp}=0$ and increases with $B_{\perp}$ regardless of the value of $B_{\parallel}$.

According to the discussions and calculations of \cite{Sch2}, all the calculations of the production rate are generalized to the case at which the magnetic fields are present if we replace $E$ in all the calculations with the following relation
\begin{align}\label{beffect}
E\to \frac{1}{\sqrt{2}}\left( E^2 -B_{\parallel}^2-B_{\perp}^2+\sqrt{\left(E^2 -B_{\parallel}^2-B_{\perp}^2\right)^2+4 E^2 B_{\parallel}^2}\right)^{1/2}.
\end{align}
Moreover, the critical electric field $E_c^B$ could be found by solving the same relation as follows
\begin{align}
E_c= \frac{1}{\sqrt{2}}\left( {E_c^B}^2 -B_{\parallel}^2-B_{\perp}^2+\sqrt{\left({E_c^B}^2 -B_{\parallel}^2-B_{\perp}^2\right)^2+4 {E_c^B}^2 B_{\parallel}^2}\right)^{1/2}.
\end{align}
A simple analysis shows that the solution to this equation is the one in Eq.\,(\ref{ecb}).
This equality is also correct for the replacements $E_c \to E_s$ and $E_c^B \to E_s^B$, which gives 
\begin{align}\label{esb}
E_s^B=E_s \sqrt{1+\frac{B_{\perp}^2}{E_s^2+B_{\parallel}^2}}.
\end{align}

\begin{figure}[h]
\begin{center}
\includegraphics[width=6.8cm]{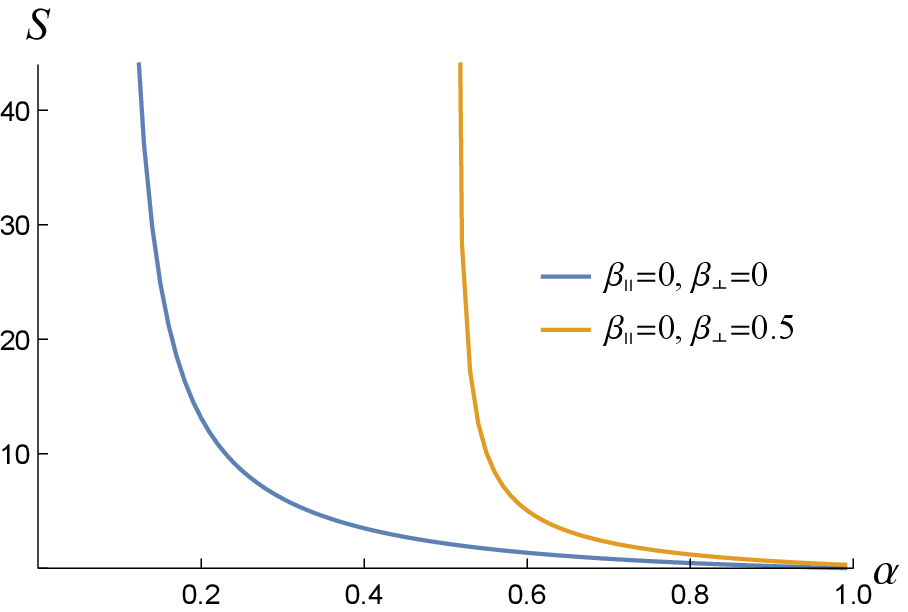}\hspace{.3cm}
\includegraphics[width=6.8cm]{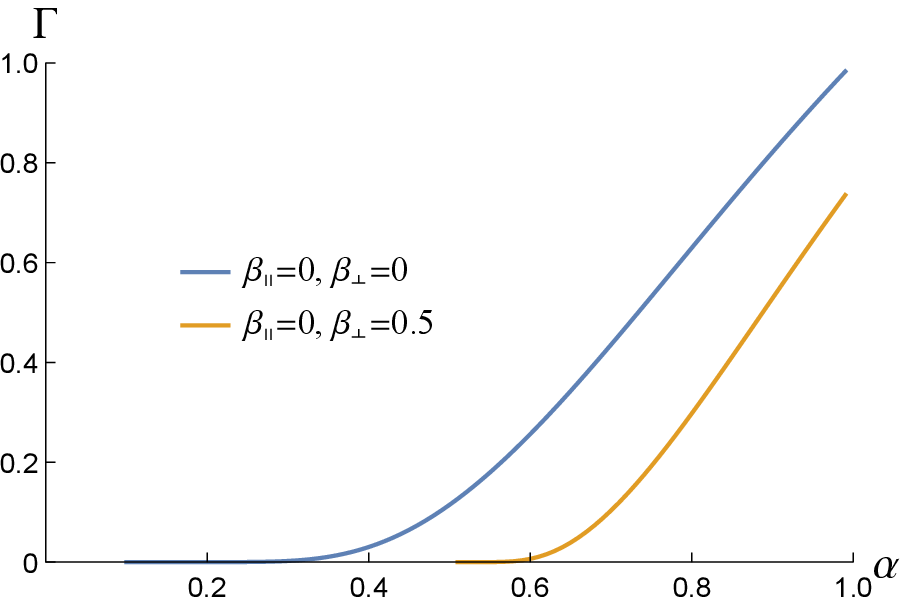}
\end{center}
\caption{\footnotesize 
The left and right graphs respectively show the effect of the parallel magnetic field on the action and the production rate in the case of the quadratic function.}
\label{decayBz}
\end{figure} 
Here, we report the results in the presence of magnetic fields.
In what follows we use the rescaled parameters $\beta_{\perp}=\frac{B_{\perp}}{E_{c}^0}$ and $\beta_{\parallel}=\frac{B_{\parallel}}{E_{c}^0}$.
Figure \ref{decayBz} shows the effect of the magnetic field perpendicular to the electric field direction.
As a result of the presence of $B_{\perp}$, the pair production starts at a larger value of the electric field as expected from Eq.\,(\ref{esb}).
Furthermore, it can be seen that $B_{\perp}$ decreases the decay rate at any given electric field, i.e., in the presence of $B_{\perp}$ we need a stronger electric field to have the same pair production rate as the case without $B_{\perp}$.

The next interesting problem is to consider what happens by turning on a magnetic field parallel to the direction of the electric field.
As can be simply observed from Eqs.\,(\ref{beffect}) and (\ref{esb}), when $B_{\perp}=0$, there remains no track of the parallel magnetic field in the calculation of the decay rate $\Gamma$.
This result is not in agreement with the results obtained in \cite{magneticdecay} where they explore the instability of confining theories influenced by external electromagnetic fields by evaluating the imaginary part of the DBI action of a probe D7-brane embedded in the bulk.
Based on their calculations, they found the same qualitative results as ours, in the case where only a magnetic field perpendicular to the electric field direction is present.
However, in the case of $(B_{\perp}=0,~B_{\parallel}\neq 0)$, they realized that although the presence of a parallel magnetic field does not alter the critical electric field $E_s$, as in our results, it definitely affects the instability of the system for $E>E_s$; the decay rate of the system is enhanced due to the presence of $B_{\parallel}$.
In fact, they concluded that the behavior of the system under the effect of the perpendicular magnetic field shows a magnetic catalysis, while the effect of the parallel magnetic field can be interpreted as an inverse magnetic catalysis.

\begin{figure}[h]
\begin{center}
\includegraphics[width=6.8cm]{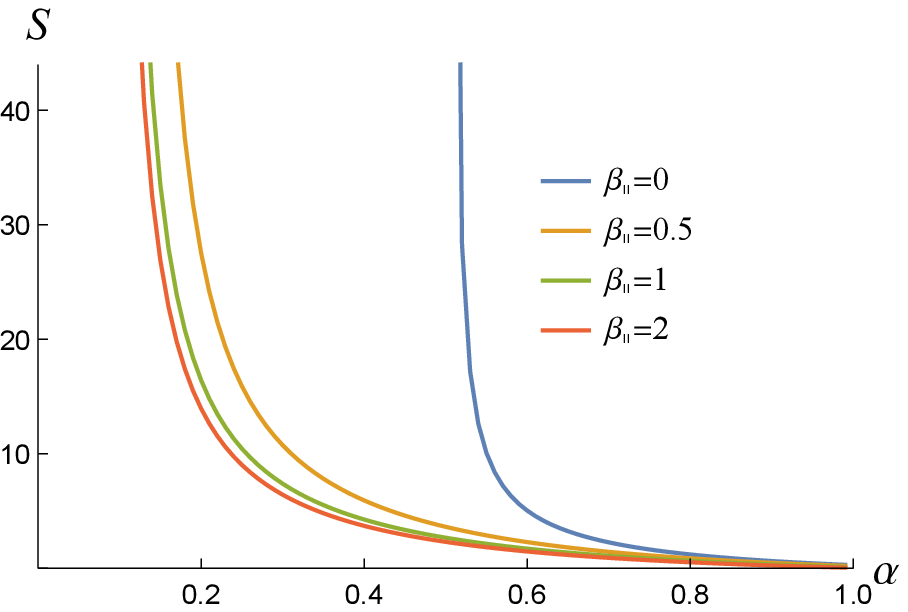}\hspace{.3cm}
\includegraphics[width=6.8cm]{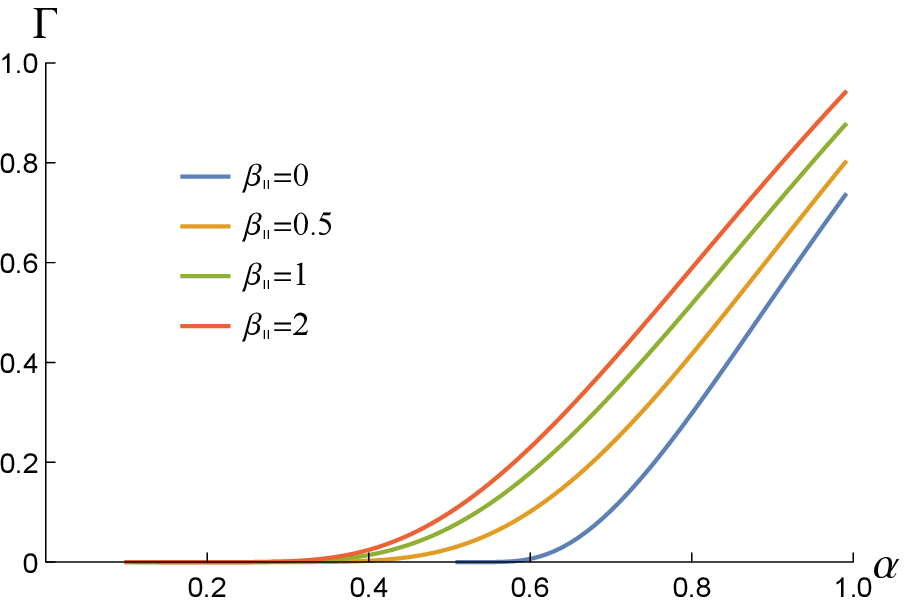}
\end{center}
\caption{\footnotesize 
The left and right graphs represent, respectively, the classical action and the decay rate as a function of $\alpha$ for $\beta_{\perp}=0.5$ and different values of $B_{\parallel}$.}
\label{decQBx}
\end{figure}

Surprisingly, here in our calculations we observe no effect from the presence of the parallel magnetic field.
However, by the use of Eqs.\,(\ref{beffect}) and (\ref{esb}), we see that the instability of the system is influenced by $B_{\parallel}$ when both the components of the magnetic field are present.
To explore the effect of $B_{\parallel}$ when there exists also a perpendicular magnetic field, the classical action and the decay rate of the theory are drawn as a function of the rescaled electric field $\alpha$ in the left and right graphs of Fig.\,\ref{decQBx}, respectively.
As can be seen from the graph of $\Gamma$, the pair production rate increases by increasing the parallel magnetic field, showing an inverse magnetic catalysis as expected from the results in \cite{magneticdecay}.

\begin{figure}[h]
\begin{center}
\includegraphics[width=6.8cm]{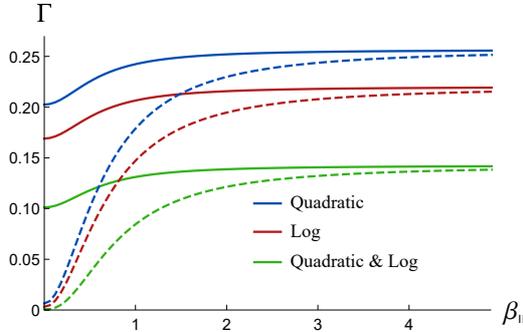}
\end{center}
\caption{\footnotesize 
The decay rate as a function of the rescaled parallel magnetic field $\beta_{\parallel}$ with $\beta_{\perp}=0.2,~0.5$ for solid and dashed lines, respectively. In all cases $\alpha=0.6$. }
\label{decay3Bx}
\end{figure}

We moreover depict the pair production rate versus the rescaled parallel magnetic field while the rescaled electric field $\alpha$ and the rescaled perpendicular magnetic field $\beta_{\perp}$ are fixed and nonzero, as indicated in the caption of Fig.\,\ref{decay3Bx}.  
In this figure the results have been drawn for three deformation functions that we are interested in, for comparison.
The results for the quadratic and logarithmic functions are closer to each other than the other function.
From this figure one can see that for a given deformation function the production rate approaches a constant value as $\beta_{\parallel}$ goes to infinity, regardless of the value of $\beta_{\perp}$.
And, the asymptotic value is the same as the value of $\Gamma$ for the given $\alpha$ in the absence of all the magnetic field components.

\begin{figure}[h]
\begin{center}
\includegraphics[width=6.8cm]{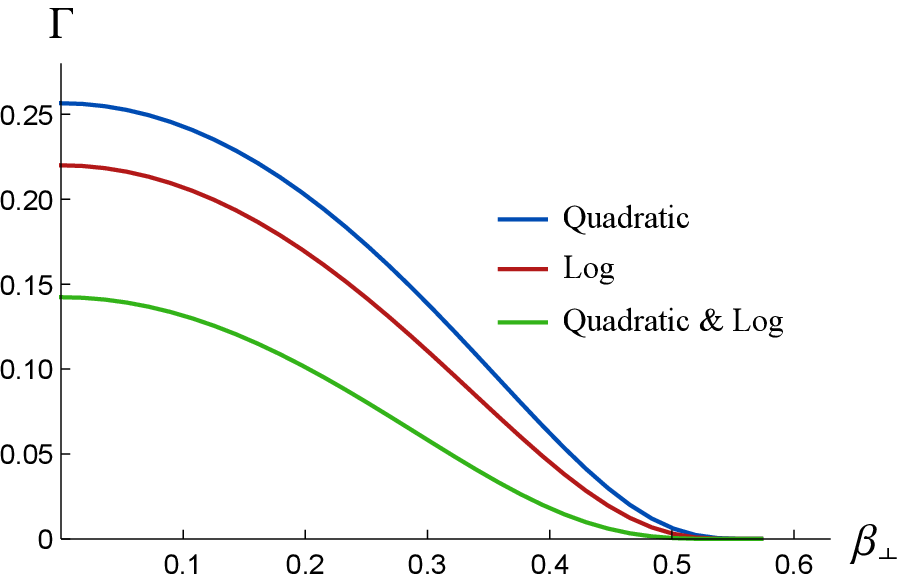}\hspace{.3cm}
\includegraphics[width=6.8cm]{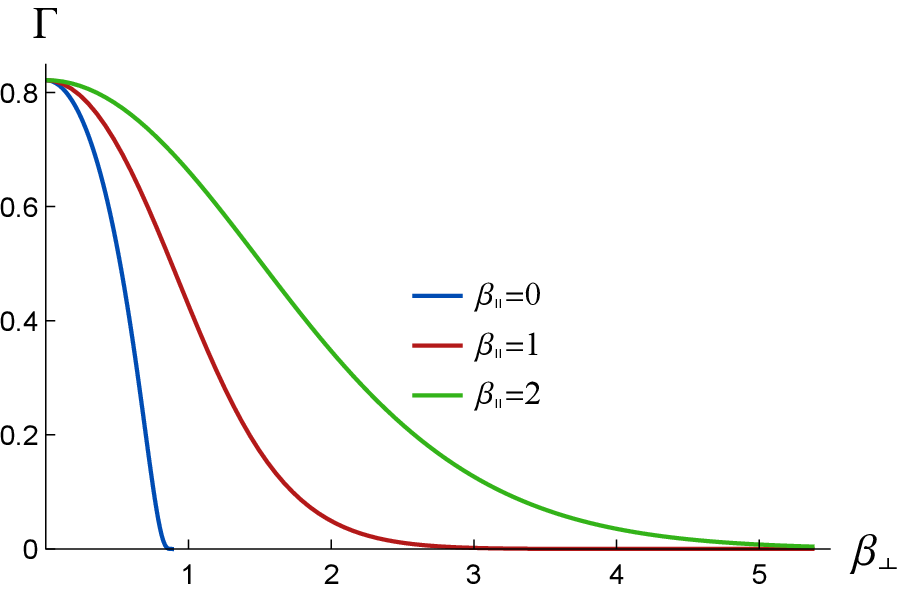}
\end{center}
\caption{\footnotesize 
Left graph: Decay rate versus the rescaled perpendicular magnetic field $\beta_{\perp}$ for $\alpha=0.6$, $\beta_{\parallel}=0$ and different deformation functions. Right graph: Decay rate versus $\beta_{\perp}$ for the case of the quadratic function with $\alpha=0.9$ and different values of $\beta_{\parallel}$.}
\label{decQBz}
\end{figure}

Another interesting result extracted from this figure can be found in $\beta_{\parallel} \to 0$.
It can be seen that in the absence of the parallel magnetic field, the production rate for three functions of our interest would be more similar when the value of $\beta_{\perp}$ increases.
In fact, the difference between the theories is removed since as $\beta_{\perp}$ becomes large enough, the production rate vanishes, as shown in the left graph of Fig.\,\ref{decQBz} for a sample value of $\alpha$.
Similar graphs, for the case of the quadratic function, are drawn in the right panel of this figure, where we can see that the presence of the parallel magnetic field postpones the decrease of the vacuum decay rate, as expected.

%************************************************************************
\section{Summary and conclusion}
The response of a confining field theory to a constant homogenous electromagnetic field has been thoroughly investigated using the gauge/gravity duality.
We have chosen the metric of the gravity side to be a deformed $\mathrm{AdS}_5$ with a deformation function containing a quadratic and a logarithmic term.
The metric has three free parameters which have been fixed by virtue of the similarity to the Cornell potential as done in \cite{quadlog}.
We have compared some of our results for three deformation functions; the quadratic function, the logarithmic function and the function with both quadratic and logarithmic terms. 
A comment is in order here.
The values we have used for the parameters only works as an approximation for our problem, since here we deal with the quarks with finite mass as required for the investigation of the Schwinger effect. 
However, in \cite{quadlog} they found these parameters using the comparison of the heavy-quark potential with the Cornell potential of heavy quarks.

The critical electric fields of the system and its total potential have been found by extremizing the action of a string attached to a D3-brane placed in the bulk near the boundary, and also the IR cutoff of the three theories, which is a generic feature of confining theories, has been obtained.
We have learnt that the potential barrier the quarks are facing with, is smaller in the case of the theory with a quadratic deformation function and the theories with only logarithmic and then with both logarithmic and quadratic functions have larger barrier potential, respectively.
%The instability of the system is suppressed due to the presence of the perpendicular magnetic field while it is enhanced by the parallel magnetic field. 

Furthermore, we have investigated the Schwinger effect for the system of our interest through the pair production rate which is calculated from the expectation value of the circular Wilson loop in the temporal-spatial plane on a probe D3-brane located at an intermediate radial position.
Holographically, this is evaluated by extremizing the NG action of a string whose world-sheet ends on the boundary of the circular Wilson loop.
We have also turned on a magnetic field to consider the effect of the simultaneous presence of electric and magnetic fields.
The aforementioned result for the three cases obtained from the potential analysis has been confirmed by exploring the results of the pair production rate in the absence of the magnetic field.

We have also observed that the presence of the perpendicular magnetic field increases the value of both of the critical electric fields, meaning that the Schwinger effect begins at a higher electric field.
It also suppresses the creation rate of the quarks for a given electric field above $E_s$.
Both of these results can be interpreted as a magnetic catalysis.
These results are in exact agreement, at least qualitatively, with the results obtained by calculating the imaginary part of the DBI action of a probe D7-brane equipped with an electromagnetic field strength \cite{magneticdecay}.
However, for the parallel magnetic field the results found by the present approach (the string approach) and the D7-brane approach are not consistent.
Here, we have found that the parallel magnetic field does not alter the critical electric fields and the instability of the system at all, while in the D7-brane approach, although $B_{\parallel}$ does not change $E_s$, it enhances the creation rate of the quarks at a given electric field.
This contradiction is probably due to the fact that the string world-sheet and DBI action are different approximations of the string theory and some particular corrections are ignored in the DBI action.
We have mentioned one other consequence of the difference between the two approaches in our previous work on the Schwinger effect \cite{LF}.

To reveal any effect of $B_{\parallel}$, we have considered the response of the system to the increase of $B_{\parallel}$ when $B_{\perp}$ is also present but have a fixed value.
In this situation, $B_{\parallel}$ starts to show off.
Its effect is the increase of the production rate consistent with the result in the D7-brane scenario.
However, the increase of $\Gamma$ due to the increase of $B_{\parallel}$ continues until at $B_{\parallel} \to \infty$, $\Gamma$ approaches an asymptotic value which is the same as the one in the absence of all components of the magnetic field.
In fact, the only effect of the increase of $B_{\parallel}$ is to compensate the decrease of $\Gamma$ due to the presence of $B_{\perp}$.
In conclusion we realize that, according to the string approach, one cannot find a higher chance of producing quarks by applying magnetic fields and the highest values for $\Gamma$ are the ones with zero $B_{\parallel}$ and $B_{\perp}$.
However, in the D7-brane approach the result is completely different; one can enhance the production rate at will and make the Schwinger effect more detectable by applying a magnetic field parallel to the electric field.
This result is important in that the Schwinger effect is too weak to be observed experimentally by the strongest electric fields that can be produced in our present laboratories and therefore any theoretical or experimental effort to make this elusive effect observable should be scrutinized.

%*********************************************************************
%\section*{Acknowledgement}
%\appendix
%\section*{Appendix A} \label{Calculation}
%\setcounter{equation}{0}
%\renewcommand{\theequation}{\Alph{section}.\arabic{equation}}

 \end{document}